\documentclass{elsart} 
 \usepackage{graphicx,natbib,amssymb} 
\journal{New Astronomy} 
\def\astrobj#1{#1}
\begin{document} 
\begin{frontmatter} 
 
\title{Near-IR spectroscopy of asteroids \astrobj{21 Lutetia}, \astrobj{89 Julia}, \astrobj{140 Siwa},
\astrobj{2181 Fogelin}, and \astrobj{5480 (1989YK8)},  potential targets for the Rosetta mission; remote observations campaign on IRTF } 
 
\author[Meudon]{Mirel Birlan\thanksref{Bucharest} \corauthref{cor} } 
\corauth[cor]{Corresponding author: Mirel.Birlan@obspm.fr} 
%\ead{Mirel.Birlan@obspm.fr}, 
\author[Meudon]{Maria Antonietta Barucci} 
\author[Meudon]{Pierre Vernazza} 
\author[Meudon]{Marcello Fulchignoni} 
\author[MIT]{Richard P. Binzel} 
\author[Hawaii]{Schelte J. Bus} 
\author[Kharkov]{Irina Belskaya\thanksref{Meudon} }
%\ead{sjb@ifa.hawaii.edu} 
\author[Padova]{Sonia Fornasier}

\address[Meudon]{Observatoire de Paris-Meudon, LESIA, 5 Place Jules Janssen, 
92195 Meudon Cedex, CNRS, France} 
\address[Bucharest]{Astronomical Institute of the Romanian Academy, Str Cutitul de Argint n 5, Bucharest 28 Romania} 
\address[MIT] {Department of Earth, Atmospheric and Planetary Sciences, Massachusetts Institute of Technology, Cambridge, MA, 02139 USA} 
\address[Hawaii]{Institute for Astronomy, 640 North A'ohoku Place, Hilo, HI 96720, USA} 
\address[Kharkov]{Astronomical Observatory of Kharkov University, Sumskaya str. 35, Kharkov 310022, Ukraine}
\address[Padova]{Astronomical Department of Padova, Vicolo dell'Osservatorio 2, 35122 Padova, Italy}

\begin{abstract} 
 
In the frame of the international campaign to observe potential target asteroids for the 
Rosetta mission, remote observations have been carried out between Observatoire de Paris, 
in Meudon-France, and the NASA Infrared Telescope Facility on Mauna Kea.
The SpeX instrument was used in the 0.8-2.5 microns spectral region, for two observing runs 
in March and June 2003.  
 
This paper presents near-IR spectra of the asteroids 21 Lutetia, 89 Julia, 
140 Siwa, 2181 Fogelin, and 5480 (1989YK8).  
 
Near-IR spectra of the asteroids 21 Lutetia and 140 Siwa are flat and featureless. 
 The spectrum of 89 Julia reveals absorption 
bands around 1 and 2 microns, which may indicate the presence of 
olivine and olivine-pyroxene mixtures and confirm the S-type designation.
 
The small main-belt asteroids 2181 Fogelin and 5480 (1989YK8) are investigated 
spectroscopically for the first time. Near-IR spectra of these asteroids also show an absorption feature around 1 micron, which could be and indicator of 
igneous/metamorphic surface of the objects; new observations in visible as well 
as thermal albedo data are necessary to draw a reliable conclusion on the surface 
mineralogy of both asteroids.

\end{abstract} 
\end{frontmatter}    
 
\section{Introduction} 
 
The observational program presented in this paper is intimately linked to the 
scientific programme of the space mission Rosetta. The Rosetta mission was 
approved in 1993 as a Cornerstone Mission within the ESA's Horizons 2000 science programme. 
The original launch date for the mission was January 2003, and the scientific targets 
were assigned to be the nucleus of comet 46P/Wirtanen and two main-belt asteroids: 140 Siwa and 4979 Otawara. 
The main scientific goals of the Rosseta mission are to investigate `in-situ' primitive objects of the Solar System and to find answers concerning the 
chemical composition of the primitive planetary nebula, thought to be `frozen' 
in the comet nucleus and the asteroid mineralogical matrix.  
 
Since 1993, several international campaigns of observations have been started in order 
to obtain a large amount of data for the targets of the Rosetta mission 
(\citet{AA2003}, \citet{AA2001}, \citet{PSS1998}, etc). 
The groundbased knowledge of these objects is essential to optimise the science return 
as well as the mission trajectory.

In January 2003 the European Space Agency decided to postpone the launch of the Rosetta, with a new launch date set to February 2004. 
This decision implied a new mission baseline: rendezvous 
with the comet P/Churyumov-Grasimenko and one or two asteroid fly-bys to be defined 
after the spacecraft interplanetary orbit insertion manoeuvre, when the total amount 
of available $\delta$v will be known.
Several fly-bys scenarios have been studied and several possible asteroid targets have 
been found. These potential targets are poorly known and for this reason systematic 
observation are needed in order to significantly improve on our knowledge of the physics 
and the mineralogy of these objects. 
 
In this paper we present the spectroscopic results obtained for five asteroids (21 Lutetia, 
89 Julia, 140 Siwa, 2181 Fogelin, and 5480 (1989YK8)), potential targets of the Rosetta mission. 
The 0.8-2.5 microns wavelength region was investigated using the SpeX instrument on IRTF-Hawaii, 
in 'remote observing' mode.

\section{The Observing Protocol} 
 
The remote observations were conducted from Meudon-France, more than 
12,000 km away from Hawaii, using several informational structures and networks. 
For the observers in Meudon, the observing hours occurred during relatively normal 
working daylight hours. Observation sessions began at 5 a.m. local time and ended 
at 5 p.m. local time.  This type of observations between Meudon and IRTF/Hawaii was 
started in 2002 (\citet{HOU2002}); since then, more than twenty nights of observations 
(in eight runs) were conducted from Meudon. The observations were realized through an 
ordinary network link, without the service quality warranty. Thus, the passband for 
our link was variable, a function of the traffic between Hawaii and Meudon.  
 
During the remote observing run, the Observatoire de Paris team had the control 
of both the instrument/guider system and  the spectrograph set-up and spectra 
acquisition. A permanent and constant audio/video link with the telescope operator  
is essential in order to administrate possible service interruptions. Two PC's 
running Linux were devoted to the  telescope/spectrograph control. The X environment 
for the telescope and instrument control was exported from Mauna Kea to Meudon 
via two secure links  (ssh tunnels). A third PC was used to keep the IP audio-video 
link open (webcam/Netmeeting at Meudon and Polycom ViewStation video-conference system on Mauna Kea). 
All software was re-initialized at the beginning of each night.
 
The communication lag was relatively small (aproximately 0.5 seconds), 
and the image refresh time was 2 seconds on the average. 
Real-time image display was performed mainly for verification purposed 
and preliminary analysis.
At the completion of each run, all files were transferred to Meudon 

The SpeX instrument was utilized in low-resolution mode for this campaign. The observations 
were made in the 0.8-2.5 microns spectral interval.  We used a 0.8 arcsec wide slit, with a 15 arcsec length and oriented North-South,
which allowed us simultaneous measurements of the object and sky. The object 
position on the slit was alternated between two locations referred to as the A and B 
positions. The seeing varied between 0.7-1.8 arcsec during the observing runs, and the 
humidity was in the 25\% - 55\% range. The automatic guiding mode of the telescope was used for spectra acquisition.

The asteroids were observed in during two observing runs: 28-30 March 2003 and 5 July 2003. 
The observed objects are summarized  in Table 1. We also observed the standard stars
\astrobj{SA 98-978}, \astrobj{SA 102-1081}, \astrobj{SA 105-56}, \astrobj{SA 107-684}, \astrobj{SA
113-276}, \astrobj{SA 115-271}, \astrobj{ HD 28099}, 
\astrobj{HD 88618}, and \astrobj{16 CyB}. Flat-fields and Argon lamp arc images were taken each night and used for data reduction. 

Our strategy was to observe all asteroids as close to the zenith as possible. Thus, we managed to observe with an 
airmass less then 1.6 for all targets except 2181 Fogelin, which we could only observe with an airmass in the 1.7-1.78 range.
Science exposures were alternated with standard stars spectra exposures, the latter taken to cover the 1-1.8 airmass range. 

In order to obtain a S/N in the 80-150 range, we needed 20 to 40 minutes of exposure time, depending on the asteroid magnitude, and counting both the effective exposure and CCD camera readout time. This exposure time is unacceptable for a 
single near-infrared spectrum due to the large variations in the atmospheric conditions (a single
near-infrared spectral exposure 
is usually no longer than 120 seconds). In order to obtain the required S/N, we obtained a number of 6 to 10 A and B exposure pairs (cycles) for both 
the asteroids -science exposures-  and the standard stars. 
This constitutes {\it one series} of observations for each object. Details the science exposures are given in Table 2.

\begin{table} 
\caption{Potential asteroid targets for the Rosetta mission. Observation date, semimajor axis, eccentricity, inclination, phase angle and
the apparent magnitude are presented for each asteroid} 
\begin{tabular}{|l|c|c|c|c|c|c|}\hline  
Object & Date &a (a.u)& e & i($\deg$)& $\Phi$($\deg$)& V(mag)\\ \hline  
21 Lutetia &March 30 2003 &  2.4347 &0.1636  &3.0645  &14.6  & 11.32  \\ 
89 Julia &March 30 2003  & 2.5519 &0.1825  &16.1437   & 13.2 &11.39\\ 
140 Siwa & March 30 2003 & 2.7365  & 0.2153  & 3.1882  & 21.2 & 13.72\\ 
2181 Fogelin & July 5 2003  & 2.5918 & 0.1177 & 13.0205   &11.2  & 16.59\\ 
5480 1989YK8 &March 30 2003   & 3.1366 & 0.087  & 6.6717   & 12.7 &16.25\\ 
 
\hline 
\end{tabular} 
 
\end{table} 
 
 \begin{table}
\caption{Exposure data for each asteroid. The columns show the mean UT value for each series, the individual time for each spectrum (Itime), the number of cycles and the mean airmass of each series.} 
\begin{tabular}{|l|c|c|c|c|}\hline 
Object & UT (h m s) & Itime(s)& Cycles& Airmass\\ \hline
21 Lutetia & 13 35 46 & 15 & 8 & 1.223       \\
21 Lutetia & 11 28 17 & 20 & 4 & 1.329       \\
21 Lutetia & 15 01 03 & 15 & 9 & 1.414      \\
21 Lutetia & 10 20 21 & 20 & 3 & 1.635  \\
89 Julia   & 10 44 44 & 40 & 6 & 1.561      \\
140 Siwa   & 07 06 07 & 120 & 8 & 1.000       \\
140 Siwa & 07 46 10 & 120 & 6 & 1.016      \\
2181 Fogelin & 12 10 35 & 2x60 & 4 & 1.769 \\ 
2181 Fogelin & 12 31 40 & 2x60 & 4 & 1.744     \\
2181 Fogelin & 12 55 17 & 2x60 & 4 & 1.740     \\
5480 1989YK8 & 12 41 01 & 120  & 6 & 1.151      \\
5480 1989YK8 & 14 36 01 & 120  & 6 & 1.230     \\
\hline 
\end{tabular} 
\end{table} 

\section{Data Reduction and Results} 
 
The major points of our reduction procedure are classic for the near-infrared spectroscopy. 
We started by combining and normalizing flat-fields for each observing night. The resulting flats 
were later used for the reduction of spectra of both the asteroids and the standard stars. In 
order to minimize the atmospheric and telescope influence and to eliminate the influence 
of electronic bias level and the dark current, we subtracted the B position spectra from 
the A position spectra for each pair of exposures (cycle), in the assumption of quasi-homogeneous 
sky background during A plus B exposure pair. The result of the subtraction was flat fielded. For each object, we median 
combined the result of all cycles in each observing series. This technique produces one positive and one negative 
spectrum on the same image.  Next step was the construction of an accurate spatial profile 
for the extraction of the spectra. The final step was the wavelength calibration.

The Spextool package (the description of the procedures are presented in \citet{PASP}) allowed 
us to perform following steps for both the asteroids and the standard stars: 
  global flat-field and arc construction, possible non-linearity correction, addition of spectra of the same object, spatial profile determination 
of the spectrum in the image, aperture location, extraction of the spectrum, wavelength 
calibration and cleaning the spectra. The results are saved in both FITS and 
ASCII formats, as used by several image processing packages (like IDL, MIDAS, and IRAF), respectively  dedicated plot/graphics software (Easyplot, 
Origin, Dataplot, etc).  

The next step in the process of data reductions was the calculation of the extinction 
coefficients. The solar analogs spectra were used to find the correspondent extinction 
coefficient for each wavelength. A "superstar" was created by summing appropriately weighted 
contributions of the standard stars. The resulting "superstar", corrected for atmospheric extinction, was used to obtain the spectral reflectance of asteroids at different 
airmasses. Careful choice of the "superstar", combined with several tests of shift between the spectrum 
of the asteroid and that of the superstar are very important in order to minimize the 
influence of the terrestrial atmosphere in the 1.4 and 1.9 microns spectral regions. 
In order to make more readable our spectra, we choose to eliminate these spectral regions.
Moreover, the elimination of these spectral regions does not affect the conclusions of this
article.

\subsection{21 Lutetia} 
With an estimated diameter of 95.5 $\pm$ 4.1 km - for an IRAS 
albedo of  0.221  $\pm$ 0.020(\citet{Tedesco1992}) - ,  21 Lutetia belongs to the large asteroid class (diameter $\ge$ 100 km). 
Its synodic period has been computed from several lightcurve analysis and the value is 8.17 $\pm$ 0.01 hours  (\citet{AA1984}).  
Color analysis of the ECAS data (\citet{Icarus1985}) designates the asteroid 21 Lutetia to 
be part of the X complex. Also, global ECAS and IRAS thermal albedo data analysis assign Lutetia to the M taxonomic type 
(\citet{Icarus1987}, \citet{ACM1989}). The M type asteroids are considered to be part of  
the core of differentiated asteroids and the parent bodies of metallic meteorites. On the basis 
of SMASS II spectroscopic data, Bus \& Binzel(2002) proposed the new taxonomic class 
$Xk$ for the asteroid 21 Lutetia.    
 
  21 Lutetia was observed on UT 2003 March 29 for the entire night. 
Four series of exposures resulted in four IR spectra with S/N in the 90-140 range.
In order to detect possible surface spectral features variations, our observations  covered 65\% of the synodic period. 
These spectra are presented in Figure 1. However, a check of the physical ephemeris of 
21 Lutetia revealed a close pole-on geometry of the asteroid, so our spectra are most probaly dominated by the contribution of the same surface features on 21 Lutetia.
There are no significant spectral features in the spectral region 0.8-2.5 micron, and the slightly increasing slope varies around 0.3\%. 

The lack of features in the spectrum of 21 Lutetia makes its
taxonomic and compositional interpretation difficult.  While
the high IRAS albedo value of 0.221 leads to the previous M-type
classification, we note there is an ambiguity in determining
the albedo of Lutetia.  An alternate low albedo value of 0.09
has been reported by Zellner et al. (1977) from polarimetric
measurements and groundbased radiometric measurements.

In Figure 2 we compare our measurements of Lutetia with others
published in the literature.  We found a good match with both
SMASS IR data (spectral interval 0.8-1.6 micron) and the
52 Color Asteroid Survey (\citet{Bell1988}) spectrophotometric data.
We also compare in Figure 2 all of these Lutetia results with reflectance spectra
of meteorites from Gaffey (1976).  Our goal is to find which
meteorites may be most analogous with Lutetia in terms of near-infrared 
spectral properties.  We note the spectrum of Lutetia
is most qualitatively similar to the spectrum of the Vigarano
meteorite, a CV3 carbonaceous chondrite, while being quite a
poor match to the class IIIA iron meteorite Chulafinee.
The purpose of our comparison is to note the difficulty of
making a non-ambiguous interpretation of the composition of 21 Lutetia.

In Figure 3 we plot the negative polarization depth versus the inversion angle (\citet{PSS1996}) for both 21 Lutetia and a sample asteroids and  meteorites. 
The asteroid 21 Lutetia has the largest inversion angle ever observed for the asteroids. 
The same peculiarly large inversion angle were found in the laboratory for the carbonaceous 
chondrites of the CV3 and CO3 samples.  These types of chondrites are characterized by a 
low carbon content and thus relatively a larger albedo when compared to other types 
(\citet{Zellner1977}).

\begin{figure} 
\begin{center} 
\includegraphics*[width=14cm]{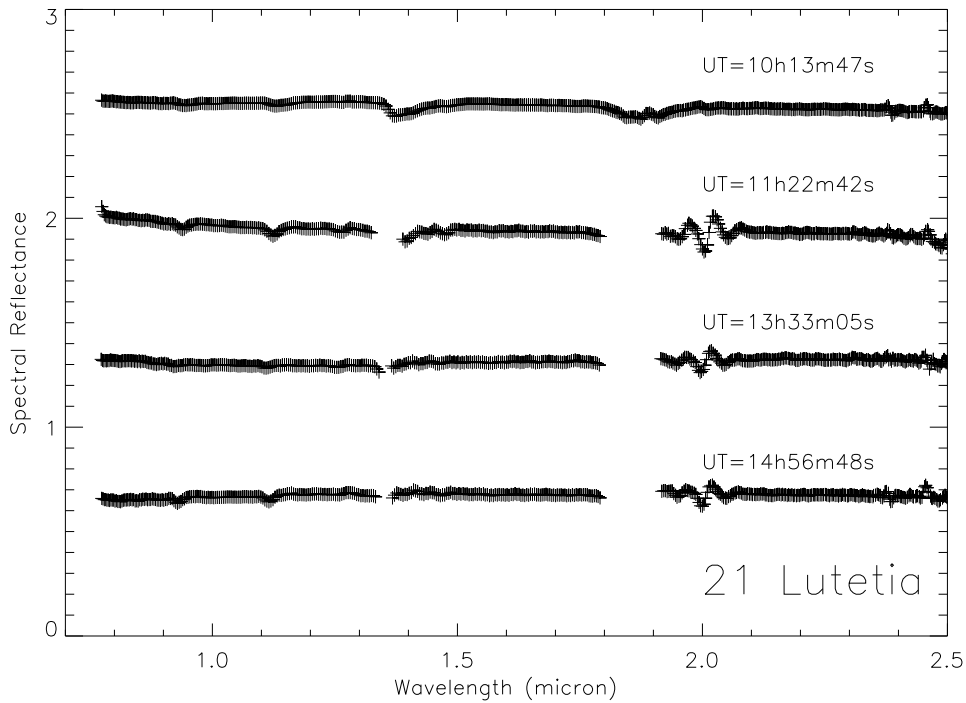} 
\end{center} 
\caption{The spectra of the asteroid 21 Lutetia. The IRTF spectra are offset for clarity and presented in chronological order. The beginning of each series of spectra(UT) was marked on the graph.} 
\end{figure}

 \begin{figure} 
\begin{center} 
\includegraphics*[width=14cm]{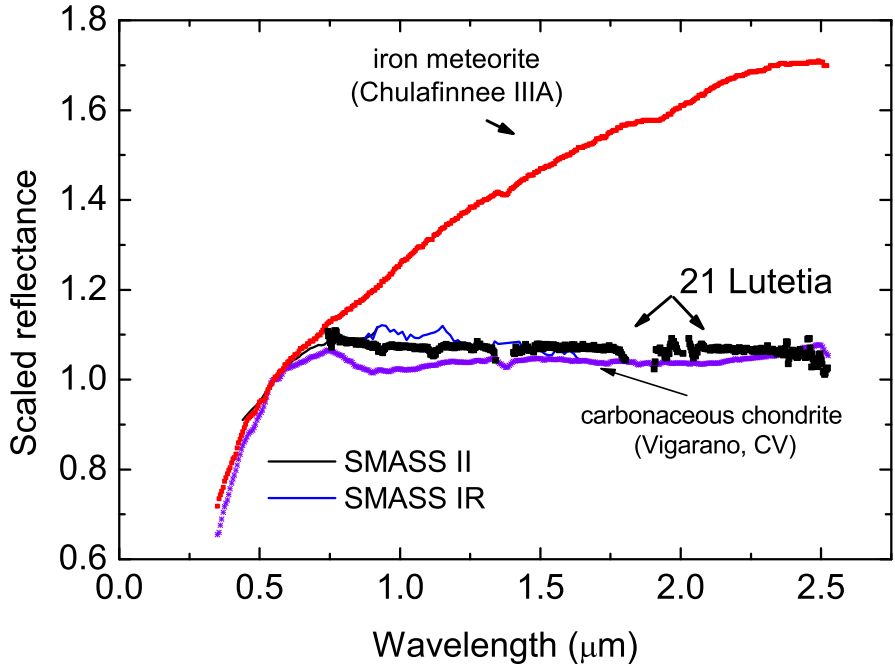} 
\end{center} 
\caption{The average spectrum of the asteroid 21 Lutetia. The SMASS II and SMASS IR data are plotted as solid lines. 
Our data is in agreement with the SMASSIR data over the 0.75-1.6 microns spectral interval. 
The comparison with Chulafinnee (iron meteorite) and Vigarano (carbonaceus chondrite) reveal 
the good match with the IR spectrum associated to carbon rich surfaces. } 

\end{figure} 

\begin{figure} 
\begin{center} 
\includegraphics*[width=14cm]{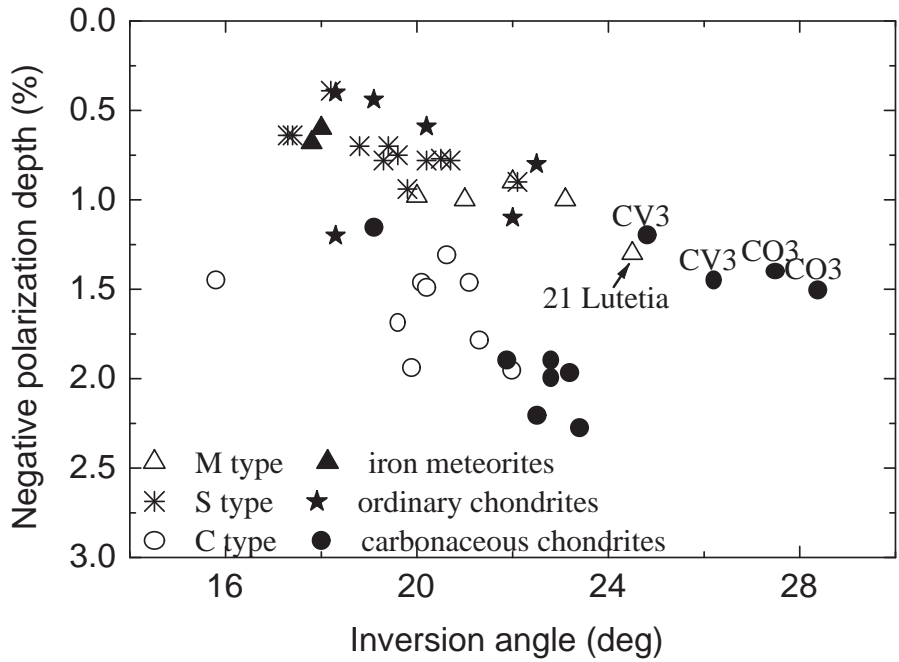} 
\end{center} 
\caption{Diagram of the negative polarization depth and the inversion angle of asteroids and 
meteorites from Belskaya \& Lagerkvist (1996). } 
\end{figure}

\subsection{89 Julia} 
 
The asteroid 89 Julia has an estimated diameter of 151.4 $\pm$ 3.1 km, for an  IRAS albedo of 0.176  
$\pm$ 0.007.  Photometry of 89 Julia yields a synodic period of 11.387 $\pm$ 0.002 hours 
(\citet{Icarus1975}). Multivariate statistics classified the asteroid as part of the S cluster (\citet{Icarus1987}, 
\citet{ACM1989}). The mineralogical classification of the S-type asteroids (\citet{Icarus1993}) based on the 
ECAS data and 52-Color spectrophotometric Survey (\citet{Bell1988}) data assigns 89 Julia (together with the 
asteroid 9 Metis)  as `ungrouped' S-asteroids. The main reason of this 'unclassification' is the long wavelength 
position of the 1-micron feature, which could be the presence of the abundant calcic clinopyroxene component 
(\citet{Icarus1993}). Based on the SMASS II data, Bus and Binzel(2002) proposed a K cluster membership (derived from 
the S cluster) for 89 Julia.
 
The asteroid spectrum was obtained for an effective integration time of 10 minutes, with a S/N ratio of 90
in both A and B beams.  This spectrum presents a significant positive slope in the region 1.1-1.5 micron, and a 
plateau in the 1.7-1.9 microns and 2.2-2.5 microns spectral regions.  In Figure 4 the IR spectrum of Julia was overlapped with the SMASS II spectrum 
in the 0.78-0.92 microns spectral region. We confirm a broad absorption band at 1 micron, typical of the silicate rich minerals. 
We also confirm a global trend of the spectral reflectance, which increases in the near-infrared.
The influence of atmospheric water absorbtion on the spectra is visible around 1.4 and 1.9 microns. The geometry of observations was 
fairly unfavourable for this asteroid (airmass = 1.56).   

In order to calculate the center of 1 micron absorbtion band, the continuum was defined
between the local maximum of spectral reflectance at 0.76 microns and the small shoulder of
the spectrum at 1.27 microns. The spectrum was then continuum subtracted and the result was
fitted with three-degree polynomial function.
 We find the center of the feature localized at 1.01 $\pm$ 0.06 micron, which is a slightly shorter wavelength than the one found by Gaffey et al (1993). 
\begin{figure} 
\begin{center} 
\includegraphics*[width=14cm]{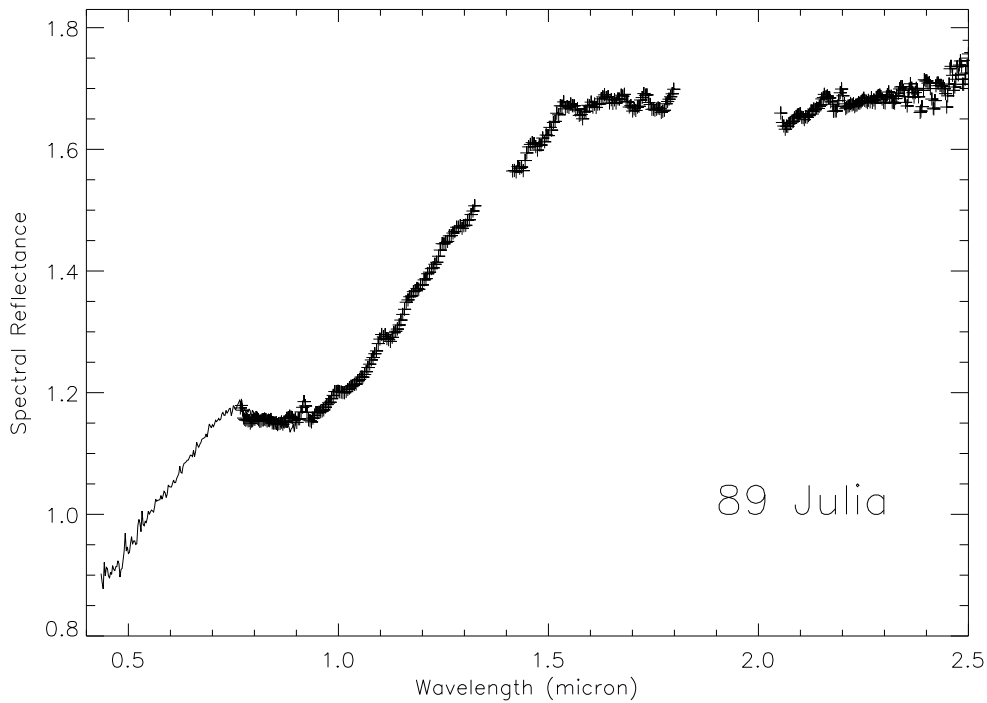} 
\end{center} 
\caption{The spectrum of the asteroid 89 Julia. The IRTF spectrum was overlapped with the SMASS II visible spectrum in the region 
0.78-0.92 micron. } 
\end{figure} 
 
\subsection{140 Siwa} 
 
140 Siwa is one of the initial targets of the Rosetta mission (departure 
in January 2003). One of the proposed scenarios (February 2003) suggests keeping this 
asteroid as a candidate for fly-by. This was our motivation to observe it.  
 
An IRAS albedo of 0.068 $\pm$ 0.004 for the asteroid 140 Siwa results in a of 109.8 $\pm$ 3.0 km diameter. 
Lightcurve analysis for two runs in 2000 (\citet{AA2001}) 
reveals a slow-rotator asteroid; the composite lightcurve presents an amplitude of 0.1 
magnitudes and its synodic period was estimated at 18.495 $\pm$ 0.005 hours.  
 
Two series of near-infrared spectra were obtained in March 30, 2003, with the ratio S/N of 50 
and 70 (obtained in both A and B beams) for the spectrum obtained at airmass 1.001 
and 1.018 respectively (Figure 5). The IR spectrum of 140 Siwa does not contain deep 
absorption features corresponding to mafic minerals. The spectrum slope is slightly positive, with a value 1\% (0.8\% and 0.7\% respectively), which confirms the slope trend 
of its spectrum (\citet{AA2001}). We confirm a typical neutral spectrum of consistent with the a C-type 
asteroid; the IRAS albedo for 140 Siwa also fits very well the average value of C taxonomic class. 
The near-IR slope of our spectra is slightly different of those of a typical C-type asteroid but this alone cannot indicate a 
different mineralogy of the surface. 

Figure 6 shows the computed a global IRTF spectrum for the asteroid 140 Siwa overlapped with SMASS II data (the ovelap is only for the  0.7-0.85 microns region). 
This composite spectrum was compared to meteorite reflectance spectra from the Gaffey database. 
There is no meteorite spectrum that fits our spectrum very well. However, the carbonaceous 
chondrite meteorites spectra are similar to our spectrum, and the asteroid spectral reflectance spans 
a value in the range of CV3 meteorite class. The nearest spectra are those of Grosnaja and Mokoia meteorites.

\begin{figure} 
\begin{center} 
\includegraphics*[width=14cm]{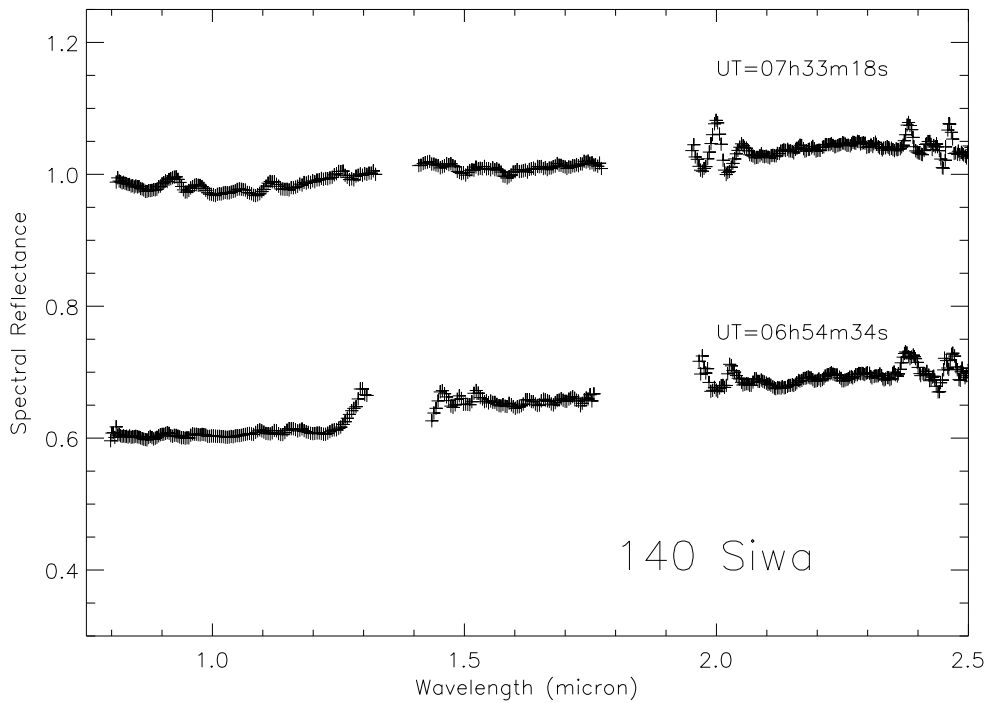} 
\end{center} 
\caption{The spectra of the asteroid  140 Siwa. The spectra were offset for clarity and the UT represents the beginning of each series of exposures. } 
\end{figure} 
 
\begin{figure} 
\begin{center} 
\includegraphics*[width=14cm]{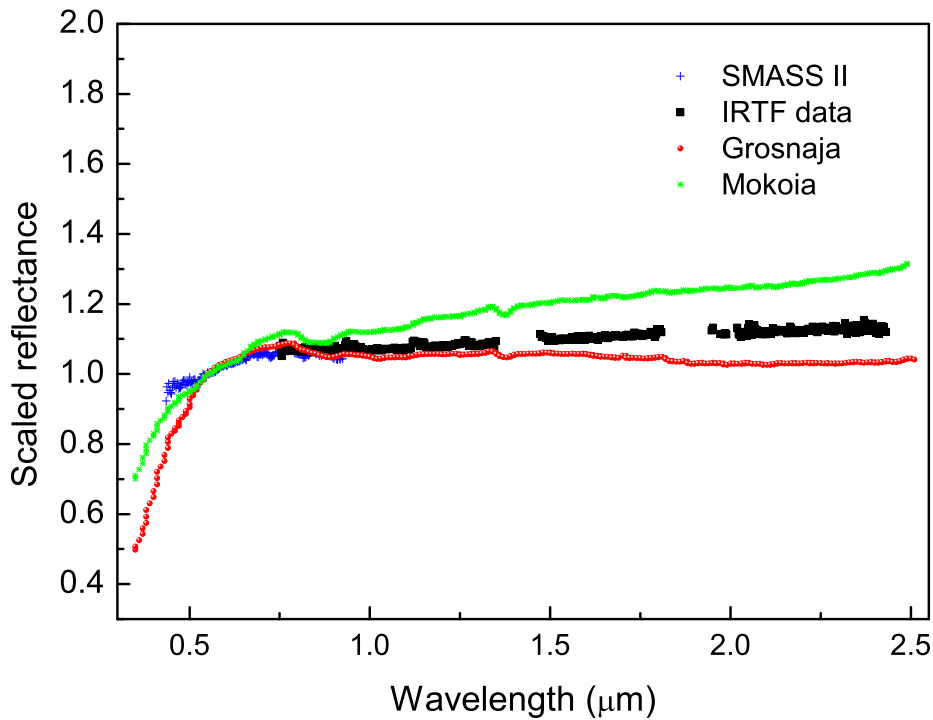} 
\end{center} 
\caption{Cumulated spectral reflectance of the asteroid 140 Siwa obtained during the IRTF run. 
The spectrum was overlapped with the SMASS II data in the 0.70-0.85 microns spectral region. 
The spectrum is similar to CV3 meteorites Grosnaja and Mokoia. The surface of 140 Siwa shows similar properties with
carbonaceous chondrite meteorites, with a reflectance trend between the curves for the two meteorites.  } 

\end{figure}

\subsection{2181 Fogelin} 
No physical data are available in the literature for the asteroid 2181 Fogelin. Thus, we can only roughly estimate its diameter. 
Following the asteroid mass distribution proposed by Kresak (1977) and the H magnitude of 
2181 Fogelin, the diameter could be in the 12-18 km range. A second estimate can be made using an empirical 
relationship between the absolute magnitude, diameter and geometric albedo (\citet{IMPS1992}):

\begin{equation}
\log{D} = 3.1236 - 0.5 \cdot \log{p_v} -0.2 \cdot H
\end{equation}

Using an IRAF albedo of 0.05-0.25 we find a 12-22 km diameter for the asteroid.
 
The observations of 2181 Fogelin were carried out in July, 5, 2003, with the ratio S/N of 80 in both A and B beams. 
The geometry of observation (airmass = 1.7) was unfavourable. However, the final spectrum is the average of 32 individual spectra, which can alleviate the high airmass problem. 
The IR spectrum presented in Figure 7 shows the large broad band absorbtion around 1 micron, typical of mafic minerals. 
The slightly positive slope in the 1.2-1.5 microns wavelength region  doesn't allow us to classify this object as a S-type asteroid. 
Further observations in the visible region as well as thermal albedo data are necessary to make a taxonomic class assignment for this asteroid.

\begin{figure} 
\begin{center} 
\includegraphics*[width=14cm]{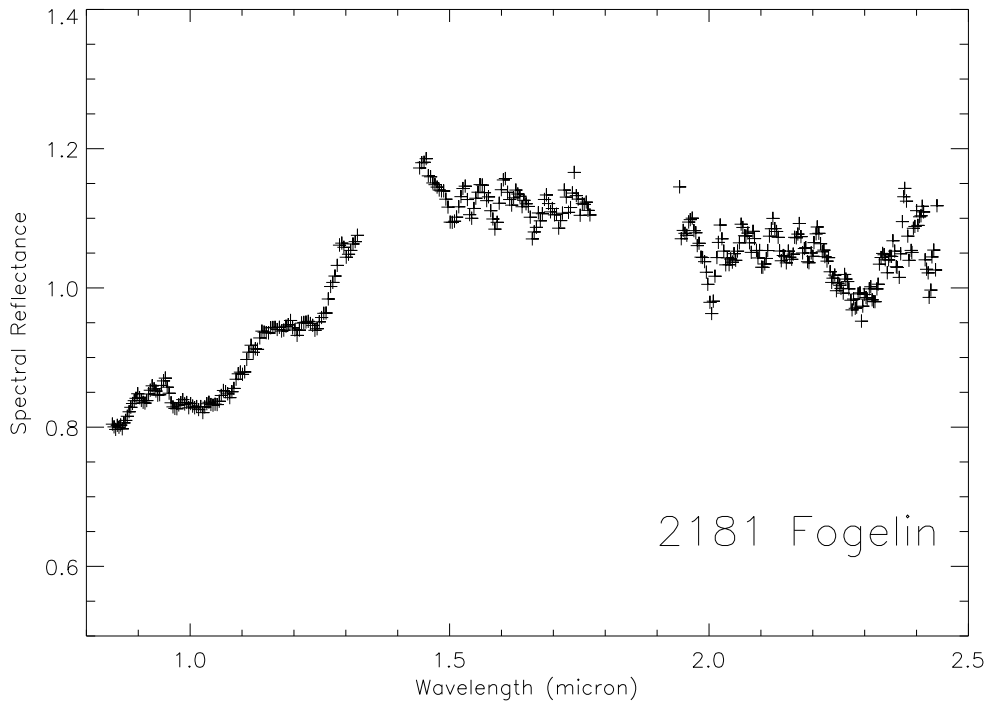} 
\end{center} 
\caption{The spectrum of the asteroid 2181 Fogelin, normalized to 1.275 microns, the maximum of J filter. } 
\end{figure}

\subsection{5480 1989YK8} 
There are no previous photometric or spectroscopic data available for the asteroid 5480 1989YK8. 
The same methods used for 2181 Fogelin have been used to compute its theoretical diameter. 
The diameter spans the 13-20 km range (following Kresak (1977)) and 13-31 km following the diameter 
empirical formula (1), used with IRAF albedos in the 0.05-0.25 range.

The asteroid was observed on March, 30, 2003, and the final spectrum S/N ratio is around 100 for the A and B positions in the slit. 
The analysis of the 0.8-2.3 microns region reveals an absorption band around 1 micron (Figure 8). 
The global tendency in near-infrared could justify a S-type asteroid classification; however, a visible spectrum and thermal albedo data are necessary for a taxonomical assignment.  
 
\begin{figure} 
\begin{center} 
\includegraphics*[width=14cm]{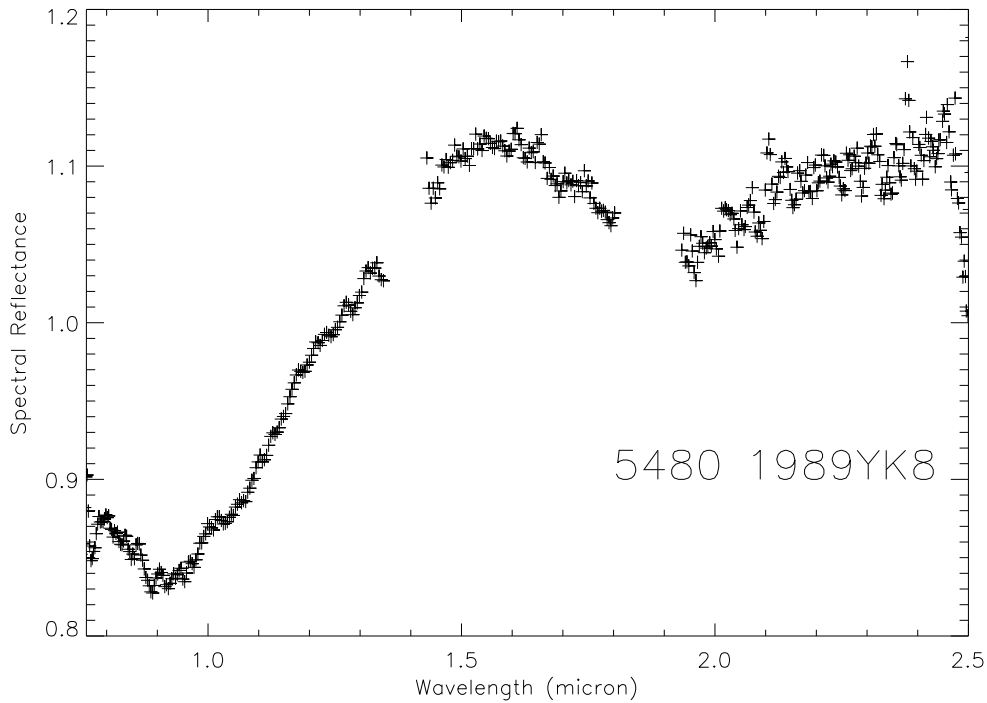} 
\end{center} 
\caption{The spectrum of the asteroid 5480 1989YK8, normalized to 1.275 microns, the maximum of J filter. } 
\end{figure}

\section{Conclusions}

Five asteroids potential targets for the Rosetta mission have been observed in the 0.8-2.5 microns spectral range: 
21 Lutetia, 89 Julia, 140 Siwa, 2181 Fogelin and 5480 (1989YK8). We confirm the flat near-IR 
spectrum of 21 Lutetia. The near-infrared spectrum of
89 Julia contains features confirming the initial idea that it belongs to the `heated' bodies, possible 
to the S or K taxonomic class. A value 1.01 $\pm$ 0.06 microns was found for the center of the Band I.   
The asteroid 140 Siwa spectrum presents a slightly positive slope, and there are no major 
mineralogical signatures; we confirm the assignament of the asteroid as an 'unheated' 
asteroid.     
 
The small asteroids 2181 Fogelin and 5480 (1989YK8) were observed spectroscopically for the 
first time. Our data shows the  presence of a large (and weak) absorption band 
around 1 micron for both asteroids; new observations in visible and infrared are necessary to draw a
reliable conclusion concerning their surface mineralogy. 

The remote observations between IRTF and Observatoire de Paris-Meudon proved to 
be a robust and handy observing technique. It offered full access to the command line of the spectrograph and to several telescope controls(focus, tracking,...). 
We consider our observing program entirely accomplished without any discernable difference in the typical amount of spectra obtained in remote 
observations mode versus the local observing mode. 
 
\section{Acknowledgements} 
 
The authors are grateful to Paul Hardersen for useful comments that improved our 
article. We are indebted to Tony Denault who answered many questions prior to our runs about 
the Polycom system and exporting of X-windows for SpeX and guiding system, to Miranda 
Hawarden-Ogata for the administration system support, and to Paul Sears and Bill Golish, the 
telescope operators during the remote observing runs.

\end{document}